\documentclass[aps,prl,twocolumn,floatfix, superscriptaddress]{revtex4-1}
\usepackage{times}
\usepackage{epsfig}
\usepackage{amsmath}
\usepackage{amssymb}
\usepackage{graphicx}
\usepackage{bbold}
\usepackage{hyperref}
\usepackage[switch]{lineno}
\usepackage{pifont}
\usepackage{bbding}
\usepackage{array}
\usepackage{multirow}
\hypersetup{colorlinks=true,citecolor=red,linkcolor=blue}

\newcommand{\tsh}{t_\mathrm{h}}

\newcommand{\hatj}{\hat{j}}
\newcommand{\hl}[1] {{\color{red} {#1}}}

\begin{document}
\title{A new pathway to impact ionization in a photo-excited 
one-dimensional ionic Hubbard model}
\author{Zhenyu Cheng}
\affiliation{College of Physics and Technology, Guangxi Normal University, Guilin, Guangxi 541004, China}
\author{Li Yang}
\affiliation{College of Physics and Technology, Guangxi Normal University, Guilin, Guangxi 541004, China}
\author{Xiang Hu}
\email{X. Hu: xianghu@gxnu.edu.cn}
\affiliation{College of Physics and Technology, Guangxi Normal University, Guilin, Guangxi 541004, China}
\author{Hantao Lu}
\email{H. Lu: luht@lzu.edu.cn}
\affiliation{Lanzhou Center for Theoretical Physics, Key Laboratory of Theoretical Physics of Gansu Province, and Key Laboratory of Quantum Theory and Applications of MoE, Lanzhou University, Lanzhou, Gansu 730000, China}
\author{Zhongbing Huang}
\affiliation{Department of Physics, Hubei University, Wuhan 430062, China}
\author{Liang Du}
\email{L. Du: liangdu@gxnu.edu.cn}
\affiliation{College of Physics and Technology, Guangxi Normal University, Guilin, Guangxi 541004, China}
\begin{abstract}
Using the time-dependent Lanczos method, we study the non-equilibrium dynamics of the half-filled one-dimensional ionic Hubbard model, deep within the Mott insulating regime, under the influence of a transient laser pulse. In equilibrium, increasing the staggered potential in the Mott regime reduces the Mott gap and broadens the Hubbard bands, creating favorable conditions for impact ionization. After laser excitation, impact ionization is observed, with its occurrence depending on both the staggered potential and the laser pump frequency. By analyzing the time evolution of the kinetic, ionic, and Coulomb interaction energies, we identify a novel mechanism for impact ionization, in which excess ionic potential energy is converted into additional double occupancy—distinct from the conventional mechanism where excess kinetic energy drives this process. 
We further show that impact ionization arises from interference between excited states driven by photon excitations of the same order.
These results present a new pathway for realizing impact ionization in strongly correlated electron systems.
\end{abstract}
\date{\today}
\maketitle

{\noindent\it Introduction} Enhancing solar cell efficiency is crucial for advancing sustainable energy and environmental conservation \cite{yuk:nrl2009,lee:rser2017,nayak:nrm2019}. For decades, the performance of single-junction solar cells has been limited by the Shockley-Queisser limit, which caps efficiency at approximately $33\%$ \cite{shockley:jap1961}. However, recent theoretical advances suggest that impact ionization in strongly correlated electronic systems could overcome this limitation, potentially enabling efficiencies exceeding $60\%$ \cite{manousakis:prb2010a, petocchi:prb2019, murakami:rmp2025}. 
In these systems, impact ionization occurs when a high-energy photon generates a charge carrier (such as a doublon or hole) with excess kinetic energy. Through strong Coulomb interactions, this carrier can de-excite, transferring its surplus energy to create an additional charge carrier, thereby amplifying the total carrier population. This process, driven by the complex electron-electron interactions in correlated materials, holds significant promise for improving the energy conversion efficiency of solar cells beyond traditional boundaries \cite{maislinger:prb2022}.

In strongly correlated electronic systems, laser exposure injects energy that can lead to the formation of doublon-holon pairs, which effectively store the absorbed energy. This energy accumulation can trigger a variety of dynamical phenomena, such as insulator-to-metal transitions \cite{Aoki:rmp2014, rincon:prb2021a,shao:prb2016a,shinjo:prb2017,werner:prb2019a}, unconventional superconductivity \cite{fausti:sci2011,huw:nm2014,kaneko:prl2019,wangy:prl2018} and  impact ionization in the high-frequency regime \cite{werner:prb2014}. 
According to the quasi-particle model, for impact ionization to occur, the laser frequency must exceed twice the charge gap of the Mott system \cite{werner:prb2014,maislinger:prb2022}. 
While this phenomenon has been observed in two-dimensional square lattices \cite{sorantin:prb2018,kauch:prb2020,Gazzaneo:prb2022,watzenbock:prb2022,maislinger:prb2022} and infinite-dimensional hyper-cubic lattices \cite{werner:prb2014},
it remains elusive in one-dimensional systems \cite{kauch:prb2020}. 
It is believed that antiferromagnetic spin fluctuations in one-dimensional materials may compete with and suppress the impact ionization process, preventing its observation \cite{kauch:prb2020}. 
The interplay between these competing effects—strong correlations, spin fluctuations, and the dynamics of photo-excited carriers—raises important questions about the limitations of impact ionization in low-dimensional systems and the conditions under which it may be induced.

In this study, we investigate the one-dimensional ionic Hubbard model (IHM), where a staggered on-site potential is introduced to the standard Hubbard model. 
The addition of this staggered potential broadens the Hubbard band bandwidth while simultaneously reducing the Mott gap. This combination of a narrower gap and an increased bandwidth can facilitate the realization of impact ionization \cite{manousakis:prb2010a,werner:prb2014}. 
The IHM can be experimentally realized in cold atomic gases using a superlattice potential \cite{pertot:prl2014, messer:prl2015a} or in condensed matter systems, such as twisted bilayer germanium selenide (GeSe) \cite{kennes:nc2020}, organic charge-transfer solids \cite{nagaosa:jpsj1986,nagaosa:jpsj1986a} and ferroelectric perovskites \cite{egami:sci1993}.

In this work, we report the observation of impact ionization in the one-dimensional IHM, evidenced by after-pulse dynamics, including an increase in double occupancy and a transfer of spectral weight within the upper Hubbard band \cite{kauch:prb2020}. 
We also propose a new pathway for impact ionization, where photo-excited charge carriers acquire excess ionic energy, which is subsequently converted into an additional doublon-holon pair. This contrasts with previous studies where excess kinetic energy is transferred into Coulomb interaction energy \cite{werner:prb2014,kauch:prb2020}. 
Further analysis of many-body excited states reveals that impact ionization arises from interference between photon-excited states of the same order. These findings offer new insights into impact ionization in strongly correlated electron systems and present new possibilities for realizing impact ionization in light-driven materials.

\indent{\it Model and Method}
 We consider the one dimensional IHM,
\begin{align}
    H =& -\tsh \sum_{i,\sigma}\left(c_{i,\sigma}^\dagger c_{i+1,\sigma}^{} + h.c.\right) + \frac{\Delta}{2} \sum_{i,\sigma}(-1)^i n_{i\sigma}  \nonumber\\
      &+\sum_i U (n_{i\uparrow} -\frac{1}{2}) (n_{i\downarrow} -\frac{1}{2}),
\label{eq:eqH}
\end{align}
where $c_{i\sigma}^\dagger$($c_{i\sigma}$) creates(annihilates) a fermionic particle at site $i$ with spin
$\sigma=\uparrow, \downarrow$ and $n_{i\sigma} = c_{i\sigma}^\dagger c_{i\sigma}^{}$ is the occupation number operator. Here
$\tsh$ is the isotropic nearest neighbour hopping amplitude, $\pm \Delta/2$ is the staggered on-site potential and $U>0$ is the local Coulomb repulsion.
In this paper, we set $\tsh =1 $ as the energy unit, with the corresponding time unit being the inverse of the energy, $\tsh^{-1}$. 
We choose the chain size to be $L=14$.
In the following, we restrict ourselves to the case of half-filling with periodic boundary conditions.
Furthermore, we assume the total magnetization in the system vanishes,
which means the number of up-spin electrons is equal to the down-spin electrons.

The sublattice-resolved partial density of states (DOS) is defined as,
\begin{align}
\rho_\nu(\omega) = \sum_{i\in\nu,\sigma} \sum_{n} 
    &|\langle n|c_{i\sigma}^\dagger|\Psi_0\rangle|^2 \delta(\omega - E_n + E_0)  \nonumber\\
   +&|\langle n|c_{i\sigma}^{     }|\Psi_0\rangle|^2 \delta(\omega + E_n - E_0)
   \label{eq:eqdos}
\end{align}
where $\nu=\mathrm{A, B}$ labels the sublattice in a single unit cell, 
$\{|n\rangle\}$ is an eigenstate of the Hamiltonian in Eq.\eqref{eq:eqH} with respective energy eigenvalue $E_n$, and $|\Psi_0\rangle$ is the ground state with energy $E_0$. The definition of the linear absorption spectrum (dynamical current-current correlation function) is \cite{luht:prl2012,okamoto:njp2019,kaneko:prl2019,Ejima:prr2022},
\begin{align}
\label{eq:labs}
    \alpha(\omega) = -\frac{1}{\pi} \mathrm{Im} \langle \Psi_0| \hatj \frac{1}{\omega - (H - E_0)} \hatj |\Psi_0\rangle,
\end{align}
with the current density operator,
\begin{align}
    \hatj = i\tsh \sum_{i \sigma} (c_{i\sigma}^\dagger c_{i+1,\sigma}^{} - c_{i+1\sigma}^\dagger c_{i,\sigma}^{}).
\end{align}

We consider a system exposed to an external laser pulse with vector potential (directed along the chain),
\begin{align}
    A(t) = A_0 \exp[-(t-t_p)^2/2t_d^2] \cos[\Omega (t-t_p)],
\end{align}
where $A_0$ is the laser intensity, $\Omega$ is the laser frequency and the laser pulse is peaked at $t_p$ with $t_d$ characterizing the duration time (pulse width). Throughout this paper, we set $t_p = 8.0, t_d = 2.0$.
The time-dependent Hamiltonian is written using the Peierls substitution,
\begin{align}
    c_{i\sigma}^\dagger c_{i+1\sigma}^{} + \mathrm{H.c.} \rightarrow e^{i A(t)}c_{i\sigma}^\dagger c_{i+1\sigma}^{} + \mathrm{H.c.}
\label{neqH}
\end{align}

The exact diagonalization method (a standard Lanczos procedure) is employed to numerically calculate the ground state of the Hamiltonian at time $t=0^-$, where the lase pulse is not yet applied to the system.
The ground state is used as an initial state for the time dependent Schr\"odinger equation $i\partial_t |\Psi(t)\rangle = H(t) |\Psi(t)\rangle$.
The time evolution is implemented step-by-step based on the time-dependent Lanczos method \cite{Park:jcp1986,Mohankumar:cpc2006,Balzer:jpcm2012,Lu:prl2012,innerberger:epjp2020},
\begin{equation}
  |\Psi(t+\delta t)\rangle \approx e^{-i H(t)\delta t} |\Psi(t)\rangle
                           \approx \sum_{l=1}^M e^{-i \epsilon_l^{} \delta t} |\Phi_l\rangle\langle \Phi_l| \Psi(t)\rangle,\nonumber
\end{equation}
where $\epsilon_l^{}$ ($\Phi_l$) are the eigenvalues (eigenvectors) of the tri-diagonal matrix generated by Lanczos iteration with $M \leq 100$. 
We set the time step size $\delta t = 0.005t_h^{-1}$ in our calculation of the time evolution.

\begin{figure}[t]
\centering
\includegraphics[angle=-0,width=0.23\textwidth]{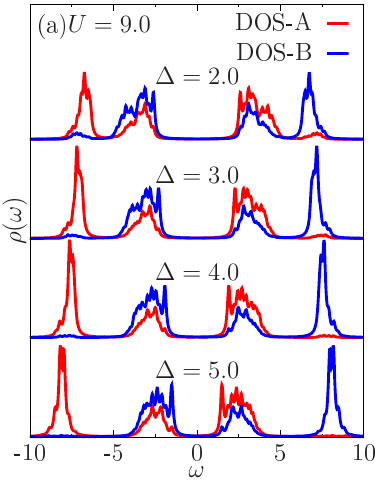}
\includegraphics[angle=-0,width=0.23\textwidth]{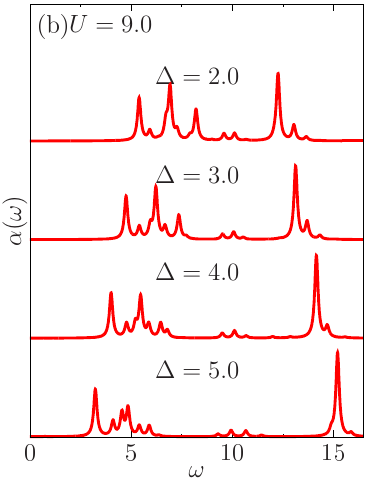}
\caption{(Color online) (a) The sublattice (A and B) resolved density of states in equilibrium calculated using Eq.\eqref{eq:eqdos}. (b) The linear absorption spectrum as a function of frequency calculated using Eq.\eqref{eq:labs}, where linear response theory applies. The broadening factor is set to $\eta = 0.1$ for both the density of states and the linear absorption spectrum. The Coulomb interaction strength is fixed at $U = 9.0 t_h$ in the ionic Hubbard model with  different staggered potential $\Delta = 2.0, 3.0, 4.0, 5.0$, respectively.}
\label{Fig:eqdosD2}
\end{figure}
\noindent{\it Linear absorption spectrum}
To pave the way for out-of-equilibrium study of the driven IHM, we study the equilibrium DOS and the linear absorption spectrum of the one dimensional IHM firstly.
Throughout this work, we focus our attention on the Hubbard superlattice deep in the Mott insulating phase \cite{fabrizio:prl1999,zhangyz:prb2003,manmana:prb2004}, where the system parameters are $U = 9.0$ and  $\Delta = 2.0, 3.0, 4.0, 5.0$.
In Fig.\ref{Fig:eqdosD2}(a), we plot the sublattice-resolved DOS (Eq.\eqref{eq:eqdos}) at zero temperature using the standard Lanczos method. 
Evidently, the particle-hole {\it like} symmetries \cite{demarco:prr2022} are observed, where the particle-hole transformation of the A-sublattice DOS is the B-sublattice DOS and each unit cell is half-filled.
The charge gap decrease monotonically with increasing staggered potential (from $5.21$ for $\Delta=2.0$ to $3.01$ for $\Delta=5.0$).  
In the Mott regime, the staggered potential splits both the lower Hubbard bands (LHB) and upper Hubbard bands (UHB) into two distinct parts for each sublattice.
The physical picture of this splitting can be understood as follows:
at the strong Coulomb interaction limit ($U \gg t_h$), the LHB is split into two parts, located at $- U/2 \pm \Delta/2$. 
The DOS at $- U/2 - \Delta/2$ is dominated by the A sublattice, while at $- U/2 + \Delta/2$, the A sublattice DOS is notably smaller than that of the B sublattice. 
Correspondingly, the UHB is the particle-hole {\it like} transformation of the LHB.

To study the effect of an external laser drive on the equilibrium system, we first study the energy absorption of the superlattice system at low laser intensity where linear response theory applies. 
In Fig.\ref{Fig:eqdosD2}(b), the linear absorption spectrum $\alpha(\omega)$ with $0.0 \leq \omega \leq 16.0$ is calculated using Eq.\eqref{eq:labs}, where Fermi's golden rule applies \cite{watzenbock:prb2022}. 
Given that impact ionization occurs when the photon energy exceeds twice the charge gap \cite{kauch:prb2020}, we focus on the high-frequency photo-excitation regime across various staggered potentials in the linear absorption spectrum. In this regime, optical excitations primarily involve transitions from the lower part of the LHB to the upper part of the UHB in the DOS.

\begin{figure}[t]
\centering
\includegraphics[angle=-0,width=0.23\textwidth]{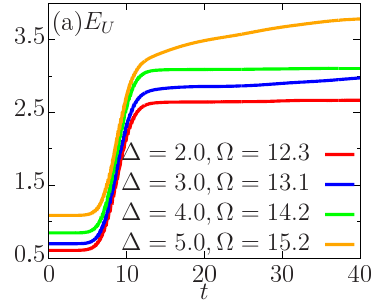}
\includegraphics[angle=-0,width=0.23\textwidth]{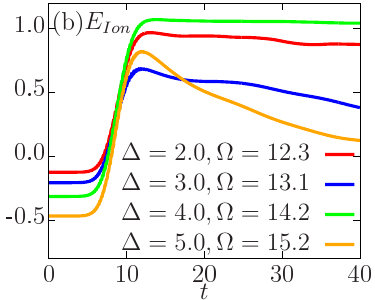}
\includegraphics[angle=-0,width=0.23\textwidth]{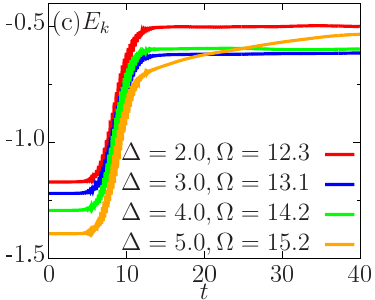}
\includegraphics[angle=-0,width=0.23\textwidth]{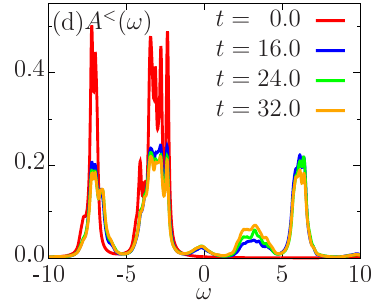}
\caption{(Color online) The time-dependent of (a) Coulomb repulsion energy, (b) ionic energy and (c) kinetic energy for different staggered potential $\Delta$ and pump frequency $\Omega$, $\Delta = 2.0, \Omega = 12.3$, $\Delta = 3.0, \Omega = 13.1$, $\Delta = 4.0, \Omega = 14.2$, $\Delta = 5.0, \Omega = 15.2$, respectively. 
(d) Non-equilibrium spectral function at various time $t=0.0, 16.0, 24.0, 32.0$ with $\Delta = 3.0$.
The Parameter of system: a lattice with $ L = 14, U = 9.0$, and laser strength $A_0 = 0.2$.}
\label{Fig:NeqEnergy}
\end{figure}

\noindent{\it Impact ionization and its dependence on $\Delta$}:
In the strong Coulomb interaction limit ($U \gg \Delta, t_h$), the Mott gap is $U - \Delta$, and the Hubbard band bandwidth is $\Delta$. As $\Delta$ increases, the gap decreases, while the bandwidth increases. Here, we increase the staggered potential strength to search for the possible impact ionization in the one dimensional IHM.
In Fig.\ref{Fig:NeqEnergy}, we plot the time dependent Coulomb, ionic and kinetic energy for $\Delta=2.0, 3.0, 4.0, 5.0$ with laser frequency at $\Omega=12.3, 13.1, 14.2, 15.2$ (determined from Fig. \ref{Fig:eqdosD2}(b)), respectively. 
At short time regime $t < 12.0$, the Coulomb , ionic and kinetic energy all increase with time. 
As the laser passes away ($t \ge t_e$), the behavior of those energy can be different, depending on the value of the staggered potential $\Delta$.

For $U=9.0, \Delta=2.0$, the Coulomb interaction and ionic energies stabilize as the laser fades away, indicating that double occupancy and ionicity remain nearly unchanged, which suggests that impact ionization is absent.
In contrast, the Coulomb interaction energy increases with time for $\Delta=3.0$ after the laser passed, which indicate that the double occupancy increase after the laser and correspondingly the impact ionization occurs \cite{werner:prb2014,kauch:prb2020}. 
By inspecting the after pulse kinetic energy in Fig.\ref{Fig:NeqEnergy}(c) of $\Delta=3.0$, the kinetic energy keeps constant after $t > 16.0$. 
Since the total energy remains constant as the laser fades, we conclude that the increase in Coulomb interaction energy (Fig.\ref{Fig:NeqEnergy}(a)) (double occupancy) is accompanied by a decrease in ionic energy (Fig.\ref{Fig:NeqEnergy}(b)). The excess ionic energy is converted into Coulomb energy through the formation of an additional doublon-holon pair, which constitute the impact ionization in the one-dimensional IHM.

On the other hand, as proposed in previous work \cite{werner:prb2014,kauch:prb2020}, besides the signal that impact ionization occur as the double 
occupancy continue to increase as the laser fade away, another evidence of impact ionization is the lesser spectral function which 
describe the occupied states, where the upper part of UHB shift its weight to lower part of UHB and 
the weight of LHB is reduced \cite{maislinger:prb2022}. 
Following the idea above, we plot the total spectral function [Eq.\eqref{Eq:neqsp} below] in Fig.\ref{Fig:NeqEnergy}(d) at 
specific times $t=0.0, 16.0, 24.0, 32.0$. The non-equilibrium spectral function are obtained by a forward Fourier transformation of the lesser Green's function \cite{werner:sd2016,kauch:prb2020,innerberger:epjp2020},
\begin{align}
      A_{ii\sigma}^<(\omega,t) = \frac{1}{\pi} \mathrm{Im} \int_0^\infty e^{i\omega t_\mathrm{rel}} G_{ii\sigma}^<(t,t+t_\mathrm{rel}) dt_\mathrm{rel},
      \label{Eq:neqsp}
\end{align}
where the time cutoff is chosen as $t_\mathrm{rel} = 80.0$. The equilibrium spectral function is shown as the non-equilibrium spectral function at time $t=0.0$ 
where there is no weight above the Fermi energy. As the system is driven out of equilibrium, the spectral weight in the LHB is transferred partially to UHB, and the UHB is split into two parts. 
Note, the non-equilibrium spectral function follows the sum rule,
\begin{align}
    \int_{-\infty}^{\infty} A^<_\nu(\omega,t) d\omega = n_\nu(t) \quad (\nu =\text{A or B}),
\end{align}
where $n_{\nu}(t)$ is the time dependent density of sublattice $\nu$. 

By comparing the after-pulse spectrum at times 
$t=16.0, 24.0, 32.0$, we observe a transfer of spectral weight from the upper part of the UHB to the lower part of the UHB. Simultaneously, the spectral weight in the LHB ($\omega \leq 0.0$) decreases. This transfer of spectral weight is indicative of a dynamic rearrangement of the electronic structure. Notably, our observations align with those reported in a $4\times 3$ square lattice study \cite{kauch:prb2020, maislinger:prb2022}, where similar spectral weight transfer was observed, accompanying the occurrence of impact ionization. This suggests that the observed spectral dynamics could be a signature of impact ionization in the system, where the creation of additional charge carriers, driven by Coulomb scattering.

Thus, with evidences from dynamical behaviors after laser pulse passed, increment of double occupancy, transfer of spectral weight from upper to lower part of UHB, we conclude that impact ionization occur in the one-dimensional IHM for $\Delta=3.0$, in a new pathway.

Continue increasing the staggered potential to $\Delta=4.0, 5.0$,  we observe that the impact ionization is absent for $\Delta=4.0$ and reappear at $\Delta=5.0$. 
This dependence on the staggered potential $\Delta$ is attributed to selection rules in optical excitation and the physical interpretation will be discussed below.

\begin{figure}[t]
\centering
\includegraphics[angle=-0,width=0.23\textwidth]{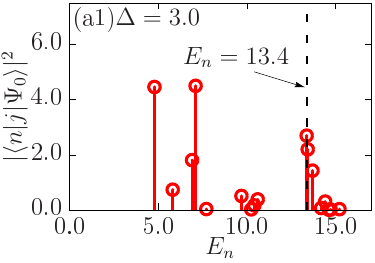}
\includegraphics[angle=-0,width=0.23\textwidth]{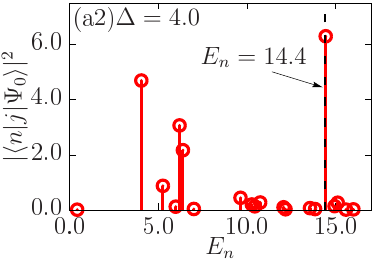}
\includegraphics[angle=-0,width=0.23\textwidth]{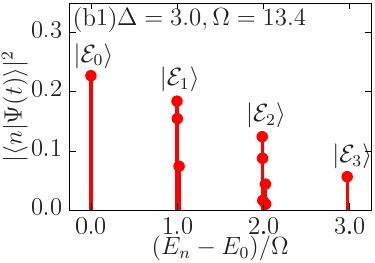}
\includegraphics[angle=-0,width=0.23\textwidth]{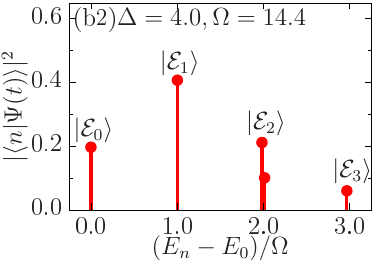}
\includegraphics[angle=-0,width=0.23\textwidth]{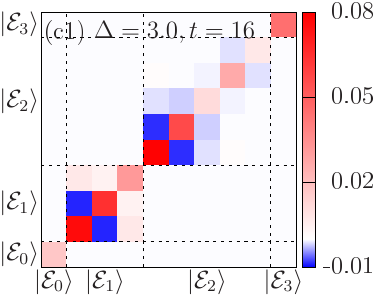}
\includegraphics[angle=-0,width=0.23\textwidth]{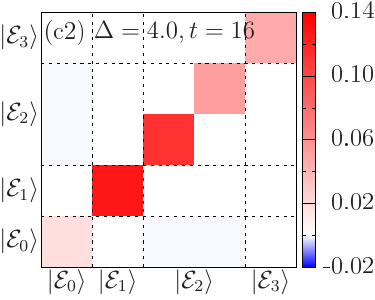}
\caption{(Color online) (a1-a2) The in equilibrium current-current correlation function $\langle jj \rangle$ as a function of frequency (Eq. \eqref{eq:jjcor}) for $\Delta=3.0, 4.0$. (b1-b2) Overlap of the after-pulse excited wave function $|\Psi(t=16)\rangle$ with the eigenvectors $|n\rangle$, $|\langle n|\Psi(t=16)\rangle|^2$. (c1-c2) Double occupancy $\mathcal{O} = \hat{D}$ elements (as shown by the colorbar) between states with probability $(|\langle n|\Psi(t=16)\rangle|^2\ge 0.01)$ in (b1-b2). 
Parameter of the lattice: $L = 8, U = 9.0$ and laser intensity $A_0 = 0.2$.}
\label{Fig:eqN8}
\end{figure}

\noindent{\it Physical mechanism of impact ionization}
To analyze the physical mechanism of impact ionization, we fully diagonalize the Hamiltonian of an 8-site chain in the IHM with $\Delta=3.0,4.0$ as a comparison, and examine the contributions from all possible many-body eigenstates. In addition to the energy conservation requirement, the even-to-odd parity condition (selection rule determined by $|\langle n|j|\Psi_0\rangle|^2$) helps identify the states that primarily influence the dynamics at the first order \cite{luht:prb2015,okamoto:njp2019,zhangyx:prb2023}. These states correspond to optically allowed excited states, providing insight into the excitation process associated with impact ionization.
For this purpose, we calculate the in equilibrium current-current correlation function,
\begin{align}
    \langle jj \rangle_n =  |\langle n|\hat{j}|\Psi_0 \rangle|^2 \delta (\omega - E_n + E_0).
    \label{eq:jjcor}
\end{align}
As shown in Figs.\ref{Fig:eqN8}[(a1) and (a2)], around the resonance frequency $\Omega=13.4$ (dashed line) for $\Delta=3.0$, there exists {\it three} states, contributing to the linear absorption spectrum $\alpha(\omega)$.
Note the Fourier transform of laser pulse to frequency space is Gaussian shape peaked with width \hl{$\tilde{\sigma} = 1/2$}, which set a resolution scale of the pump frequency $\Omega$ \cite{maislinger:prb2022}.
In contrast, for a staggered potential $\Delta=4.0$, there is only a {\it single} state around the resonant frequency $\Omega=14.3$.

Focusing on non-equilibrium dynamics with the time regime after the laser ends, where the Hamiltonian becomes time independent, and we can decompose the excited state into the eigenstates of the Hamiltonian $|E_n\rangle$.
In Figs.\ref{Fig:eqN8}[(b1) and (b2)], we plot the overlap between the excited state  $|\Psi_\mathrm{ex}(t=16.0)\rangle$ and full many-body eigenstates $|\langle E_n|\Psi_\mathrm{ex}\rangle|^2$, excluding states with a probability smaller than 0.01. 
As the lase pulse fades away, the excited wave function at any further time $t_\mathrm{e}+\Delta t$ can be written as,
\begin{align}
    |\Psi(t_\mathrm{e}+\Delta t)\rangle = \sum_n c_n(t_\mathrm{e}) e^{-iE_n\Delta t} |n\rangle,
\end{align}
with $c_n(t_\mathrm{e}) = \langle n|\Psi(t_\mathrm{e})\rangle$ and $|n\rangle$ is the eigenstate of $H(t_\mathrm{e})$ with energy $E_n$. Note $|\langle E_n|\Psi_\mathrm{ex}\rangle|^2 = |c_n(t_\mathrm{e})|^2$ is time-independent as the laser passed away.

In consistence with the equilibrium current-current correlation shown in Fig.\ref{Fig:eqN8}(a1-a2), 
there exist three energy eigenstates at the one-photon frequency ($\approx (E_n-E_0)/\Omega$) excitation which contribute sizeably to the exited state for $\Delta=3.0$, and only a single state for $\Delta=4.0$ in Fig.\ref{Fig:eqN8}(b1-b2).
The discreetness of the exited energy spectrum is attributed to charge-spin separation in one dimensional system \cite{mizuno:prb2000,itoh:prb2006,takahashi:prb2008,okamoto:njp2019}.
Taking into account that there exist only several states contribute more than $95\%$ of the excited state, the excited state can be approximately expanded as,
\begin{align}
     |\Psi(t_\mathrm{e}+\Delta t)\rangle \approx \sum_{m=0}^3 \sum_{j\in m} c_m^j (t_\mathrm{e}) e^{-i \mathcal{E}_m^j \Delta t} |\mathcal{E}_m^j\rangle,
\end{align}
where $c_m^j(t_\mathrm{e}) = \langle \mathcal{E}_m^j|\Psi(t_\mathrm{e})\rangle$ and $|\mathcal{E}_m^j\rangle$ is the eigenstate of $H(t_\mathrm{e})$ with energy $\mathcal{E}_m^j = E_0 + m\Omega + \delta_m^j$. We label a state in the manifold of $m$-photon excited states as $j\in m$, with $m \leq 3$ as a cutoff at three-photon excitation process.

With the excited wave-function above, the expectation value of any physical operator $\mathcal{O}$ is written as:
\begin{align}
   &\quad \langle\mathcal{O}(t_\mathrm{e}+\Delta t)\rangle = \langle\Psi(t_\mathrm{e}+\Delta t)|\mathcal{O}|\Psi(t_\mathrm{e}+\Delta t)\rangle \nonumber\\
    &= \sum_{n,i\in n}\sum_{m,j\in m} e^{i(\mathcal{E}_m^j - \mathcal{E}_n^i)\Delta t} c_m^{j*}(t_\mathrm{e}) c_n^i(t_\mathrm{e}) 
       \langle \mathcal{E}_m^j|\mathcal{O}|\mathcal{E}_n^i\rangle\nonumber\\
    &=  \langle\mathcal{O}(t_\mathrm{e}+\Delta t)\rangle_{m\neq n} +   \langle\mathcal{O}(t_\mathrm{e}+\Delta t)\rangle_{m=n}.
\label{eq:Delements}
\end{align}
For the first term $\langle\mathcal{O}(t_\mathrm{e}+\Delta t)\rangle_{m\neq n}$, its contribution will induce time-dependent oscillation of observable \cite{okamoto:njp2019}, 
\begin{align}
   \langle\mathcal{O}\rangle_{m\neq n} 
    \approx \sum_{m\neq n} \sum_{ij} c_m^{j*}(t_e) c_n^{i}(t_e) S_{mn}^{ji} e^{i(m-n)\Omega\Delta t},
\end{align}
with $S_{mn}^{ji} = \langle \mathcal{E}_m^j|\mathcal{O}|\mathcal{E}_n^i\rangle$.
For the second term with $m = n$, the contribution can be separated into two parts, the diagonal part ($i=j$) contribute the time-independent part of observable, 
\begin{align}
   \langle\mathcal{O}\rangle_{m=n}^{i=j} 
    = \sum_{m,j\in m} |c_m^j(t_e)|^2 S_{mm}^{jj},
\end{align}
and the off-diagonal part ($i\neq j$) responsible for the after laser dynamics, 
\begin{align}
   \langle\mathcal{O}\rangle_{m=n}^{i\neq j} 
    = \sum_{m,i\neq j} c_m^{j*}(t_e) c_m^{i}(t_e) S_{mm}^{ji}  e^{i(\delta_j-\delta_i)\Delta t},
\end{align}
where the coefficient $e^{i(\delta_j-\delta_i)\Delta t}$ induces the long time dynamics in the after pulse time regime.

Considering states with relatively large probabilities (i.e. $(|\langle n|\Psi(t=16)\rangle|^2 \ge 0.01)$) and distinguish states with the photon-excitation process (e.g., label $|\mathcal{E}_m\rangle$ means $m$-th photon excitation process), we calculate the matrix elements of total double occupancy operator $S_{mn}^{ji} = \langle \mathcal{E}_m^j|\hat{D}|\mathcal{E}_n^i\rangle$.
In Figs.\ref{Fig:eqN8}[(c1) and (c2)], the double occupancy matrix elements $S_{mn}^{ji}$ for $\Delta=3.0, 4.0$ are shown, respectively. We see that, the element between photo-excitation states with different order ($m\neq n$) are small. On the other hand, the coupling between the same order photon-excited states (off-diagonal elements of diagonal blocks) is significant for $\Delta=3.0$, while it is ignorable for $\Delta=4.0$.  These elements are responsible for the after pulse dynamics, including impact ionization.
In one word, impact ionization occurs due to the coupling between photon-excited states within the same photon process, rather than between states with different photon-excitation processes. 

In summary, we find that impact ionization occurs in the one-dimensional IHM, primarily due to interference effects within the photon-excited states of the same order.

\noindent{\it Conclusion and Discussion}:
In this work, we explore the phenomenon of impact ionization in the one-dimensional IHM, with a focus on its dependence on the staggered potential $\Delta$. 
Using evidences from the increase in double occupancy and spectral weight transfer within the UHB after the laser pulse fades, we confirm that impact ionization occurs for 
$\Delta=3.0$ and $5.0$, while it is absent for $\Delta=2.0$ and  $4.0$.
These results demonstrate that both a sufficiently strong staggered potential and the selection rule are crucial in enabling impact ionization. This finding highlights the significance of the interplay between the staggered potential and electron-electron interactions in determining the dynamics of charge carrier generation.

To further understand the underlying dynamics, we examine the time evolution of the Coulomb, ionic, and kinetic energy following the laser pulse. Our analysis reveals that the increase in the Coulomb interaction energy is directly related to a decrease in the ionic energy. This suggests that the system undergoes a redistribution of energy between different forms of excitation, with the excess ionic energy being converted into Coulomb energy. Based on this observation, we propose a new pathway for impact ionization, where excessive ionic energy triggers the process of impact ionization. This mechanism complements the previously established scenario, where impact ionization is driven by the transfer of excessive kinetic energy to Coulomb interaction energy. The involvement of both ionic and Coulomb energy in facilitating ionization provides a deeper understanding of the energy dynamics at play, highlighting the critical role of ionic energy in the impact ionization process.

To gain deeper insight into the nature behind impact ionization, we perform a full diagonalization of an 8-site chain within the IHM. This method allows us to examine the many-body eigenstates and track how physical observables evolve over time during laser excitation. A detailed analysis of the contribution of these eigenstates to the double occupancy reveals that impact ionization arises from interference between photon-excited states of the same order. Our findings shed light on the conditions required for impact ionization in strongly correlated systems and contribute to the growing body of research on controlling electron dynamics via external excitations.
\begin{acknowledgements}
We acknowledge helpful discussions with Xuedong Tian. We gratefully acknowledge funding from the National Natural Science Foundation of China (Grant No. 12464018, 12364022, 12174168).
\end{acknowledgements}
\bibliography{ihmref}
\end{document}